\documentclass[aps,twocolumn,groupedaddress]{revtex4}
\usepackage{epsfig}
\usepackage{graphicx}
\usepackage[T1]{fontenc}
\usepackage{ae}
\usepackage{color}
\usepackage[latin1]{inputenc}
\usepackage{amssymb,amsbsy,amsmath}
\usepackage{bbm}

\newtheorem{defi}{Definition}

\begin{document}

%\begin{center}\today\end{center}

%%%%%%%%%%%%%%%%%%%%%%%%%%%%%%%%%%%%%%%%%%%%%%%%%%%%%%%%%%%%%%%%%%%%%%%%%%%%%

\title[Theory of ground states V]
{Theory of ground states for classical Heisenberg spin
systems V}

\author{Heinz-J\"urgen Schmidt$^1$ and Wojciech Florek$^2$
}
\address{$^1$  Universit\"at Osnabr\"uck,
Fachbereich Physik,
 D - 49069 Osnabr\"uck, Germany\\
$^2$  Adam Mickiewicz University, Faculty of Physics,
ul.~Uniwersytetu Pozna\'{n}skiego 2, 61-614 Pozna\'{n}, Poland}

%\tableofcontents

\begin{abstract}
We formulate part V of a rigorous theory of ground states for classical, finite, Heisenberg spin systems.
After recapitulating the central results of the parts I - IV previously published we extend the theory to
the case where an involutary symmetry is present and the ground states can be distinguished according to their degree
of mixing components of different parity. The theory is illustrated by a couple of examples of increasing complexity.
\end{abstract}

\maketitle

%%%%%%%%%%%%%%%%%%%%%%%%%%%%%%%%%%%%%%%%%%%%%%%%%%%%%%%%%%%%%%%%%%%%%%%%%%%%%%%%%%%%%%%%%%%%%%%%%%%%%%%%%%%%%%%%%%%%%%%%%%%%%%%%%%%%%%%%%%
\section{Introduction}\label{sec:I}
%%%%%%%%%%%%%%%%%%%%%%%%%%%%%%%%%%%%%%%%%%%%%%%%%%%%%%%%%%%%%%%%%%%%%%%%%%%%%%%%%%%%%%%%%%%%%%%%%%%%%%%%%%%%%%%%%%%%%%%%%%%%%%%%%%%%%%%%%%

The ground state of a spin system and its energy represent valuable information, e.~g., about its low temperature
behaviour. Most research approaches deal with quantum systems, but also the classical limit has found some interest and applications,
see, e.~g., \cite{AL03} - \cite{Setal20}. For classical Heisenberg systems, including Hamiltonians with a Zeeman term due to an external
magnetic field, a rigorous theory has been recently established \cite{SL03} - \cite{S17d} that yields, in principle, all ground states.
However, two restrictions must be made: (1) the dimension $M$ of the ground states found by the theory is {\em per se} not
confined to the physical case of $M\le 3$, and (2) analytical solutions will only be possible for special couplings or small
numbers $N$ of spins. A first application of this theory to frustrated systems with wheel geometry has been given in \cite{FM19} and \cite{FKM19}.

The purpose of the present paper is to give a concise review of the central results of \cite{SL03} - \cite{S17d}
and to provide a couple of examples of increasing complexity thereby exploring the above mentioned limits of analytical treatment.
Moreover, we will cover another aspect of the ground state problem connecting the geometry of eigenvalue varieties with certain rules
of avoided level crossing known from quantum mechanics. This leads to a novel distinction between ``isolated" and ``cooperative"
ground states w.~r.~t. some symmetry $\Pi$.

To clarify the latter remarks let us recall that the theory outlined in  \cite{SL03} - \cite{S17d}, in a sense,
reduces the ground state problem to an eigenvalue problem of the matrix ${\mathbbm J}({\boldsymbol\lambda})$
of coupling coefficients of the spin system. This matrix additionally contains in its diagonal certain unknown numbers,
symbolized by ${\boldsymbol\lambda}$, that are essentially Lagrange parameters due to the constraint of the classical
spin vectors being of unit length. Hence the eigenvalues of ${\mathbbm J}({\boldsymbol\lambda})$, especially the lowest one
$j_{min}({\boldsymbol\lambda})$, should not be viewed simply as numbers but as {\em functions}. Their graphs will be called
{\em eigenvalue varieties} since they are typically not completely smooth but contain singular points or subsets.
The corresponding eigenspace (function) will be denoted by ${\mathcal W}_{min}({\boldsymbol\lambda})$.

One central result of \cite{S17a} is that the eigenvalue function $j_{min}({\boldsymbol\lambda})$ assumes its
maximum $\hat{\jmath}=j_{min}(\hat{\boldsymbol\lambda})$ at a unique point $\hat{\boldsymbol\lambda}$
and that the ground states are linear combinations of vectors from ${\mathcal W}_{min}(\hat{\boldsymbol\lambda})$
in a sense to be made more precise in Section \ref{sec:T}, see Eq.~(\ref{T7}).
The dimension of the ground states is essentially the dimension of ${\mathcal W}_{min}(\hat{\boldsymbol\lambda})$, i.~e.,
the degeneracy of the eigenvalue $j_{min}(\hat{\boldsymbol\lambda})$. This result makes it plausible
that the occurrence of two-dimensional (coplanar) ground states is connected to the rule of avoided level crossing.
Moreover, since the crossing of levels belonging to different symmetry sectors is allowed, the presence of a
symmetry leads to different types of ground states. In this paper we will concentrate on the simplest case
where we have a certain involutory symmetry $\Pi$, i.~e., satisfying $\Pi^2={\mathbbm 1}$, and accordingly
eigenspaces of different parity $\pm 1$. Then a coplanar ground state will be composed of vectors either
of the same parity (``isolated case" subject to avoided level crossing) or of different parity
(``cooperative case" subject to symmetry-allowed level crossing).

After recapitulating, in Section \ref{sec:T},  the general theory including the novel aspects sketched above,
we will, in Section \ref{sec:E}, consider three examples. All examples possess a symmetry $\Pi$ of the kind
described above, a variable bond parameter $\alpha$, and show a phase transition between a one-dimensional (collinear) and coplanar
ground states at a critical value of $\alpha=\alpha_c$. The isosceles triangle, Subsection \ref{sec:ET}, and the
square with a diagonal bond, Subsection \ref{sec:ES}, have only coplanar ground states of the cooperative type.
The last example in Subsection \ref{sec:EC} is an almost regular cube with two variable bonds of equal strength $\alpha$.
Here we observe two additional phase transitions between isolated and cooperative coplanar ground states.
We close with a Summary and Outlook in Section \ref{sec:SO}.

%%%%%%%%%%%%%%%%%%%%%%%%%%%%%%%%%%%%%%%%%%%%%%%%%%%%%%%%%%%%%%%%%%%%%%%%%%%%%%%%%%%%%%%%%%%%%%%%%%%%%%%%%%%%%%%%%%%%%%%%%%%%%%%%%%%%%%%%%%
\section{Theory}\label{sec:T}
%%%%%%%%%%%%%%%%%%%%%%%%%%%%%%%%%%%%%%%%%%%%%%%%%%%%%%%%%%%%%%%%%%%%%%%%%%%%%%%%%%%%%%%%%%%%%%%%%%%%%%%%%%%%%%%%%%%%%%%%%%%%%%%%%%%%%%%%%%

We will shortly recapitulate the essential results of \cite{S17a}-\cite{S17d} in a form adapted to the present purposes.
Let ${\mathbf s}_\mu,\; \mu=1,\ldots,N,$ denote $N$ classical spin vectors of unit length, written as the rows of
an $N\times M$-matrix ${\mathbf s}$ where $M=1,2,3$ is the dimension of the spin vectors. The energy of this system
will be written in the form
\begin{equation}\label{T1}
 H({\mathbf s})=\frac{1}{2}\sum_{\mu,\nu=1}^{N} J_{\mu\nu} {\mathbf s}_\mu\cdot {\mathbf s}_\nu,
\end{equation}
where the $J_{\mu\nu}$ are the entries of a symmetric, real $N\times N$-matrix ${\mathbbm J}$ with vanishing diagonal
elements. In contrast to \cite{S17a}-\cite{S17d} the factor $\frac{1}{2}$ is introduced for convenience.
A {\em ground state} is a spin configuration ${\mathbf s}$ minimizing the energy $ H({\mathbf s})$. If we fix
all vectors ${\mathbf s}_\nu$ of a ground state except a particular one ${\mathbf s}_\mu$, the latter has to minimize
the term
\begin{equation}\label{T2}
 H_\mu\equiv {\mathbf s}_\mu\cdot\left( \sum_{\nu=1}^N J_{\mu\nu} {\mathbf s}_\nu\right)
 \;.
\end{equation}
Hence ${\mathbf s}_\mu$ must be a unit vector opposite to the bracket in (\ref{T2}) and thus has to satisfy
\begin{equation}\label{T3}
  -\kappa_\mu\,{\mathbf s}_\mu=\sum_{\nu=1}^N J_{\mu\nu} {\mathbf s}_\nu
  \;,
\end{equation}
with Lagrange parameters $\kappa_\mu\ge 0$. Upon defining
\begin{equation}\label{T4}
  \overline{\kappa}\equiv\frac{1}{N}\sum_{\mu=1}^{N}\kappa_\mu, \quad \mbox{and } \lambda_\mu\equiv\kappa_\mu-\overline{\kappa}
  \;,
\end{equation}
such that
\begin{equation}\label{T4a}
 \sum_{\mu=1}^{N}\lambda_\mu=0
  \;,
\end{equation}
we may rewrite (\ref{T3}) in the form of an eigenvalue equation
\begin{equation}\label{T5}
  \sum_{\nu=1}^{N}{\mathbbm J}_{\mu\nu}({\boldsymbol\lambda})\,{\mathbf s}_\nu\equiv
   \sum_{\nu=1}^{N}\left( J_{\mu\nu}+\delta_{\mu\nu}\lambda_\nu\right){\mathbf s}_\nu=
   -\overline{\kappa}\,{\mathbf s}_\mu
   \;.
\end{equation}
Here we have introduced the {\em dressed $J$-matrix} ${\mathbbm J}({\boldsymbol\lambda})$ with vanishing trace considered as a function
of the vector ${\boldsymbol\lambda}=(\lambda_1,\ldots,\lambda_N)$ of ``gauge parameters".

We denote by $j_{min}({\boldsymbol\lambda})$ the lowest eigenvalues of ${\mathbbm J}({\boldsymbol\lambda})$ and by
${\mathcal W}_{min}({\boldsymbol\lambda})$ the corresponding eigenspace. It can be shown \cite{S17a} that
the graph of the function
$j_{min}({\boldsymbol\lambda})$, the ``eigenvalue variety",
has a maximum, denoted by $\hat{\jmath}$, that is assumed at a uniquely determined point $\hat{\boldsymbol\lambda}$ such that
\begin{equation}\label{T6}
 E_{min}=\frac{1}{2}\,N \, \hat{\jmath}
\end{equation}
is the ground state energy and that the ground state configuration ${\mathbf s}$ can be obtained as a linear combination
of the corresponding eigenvectors of ${\mathbbm J}(\hat{\boldsymbol\lambda})$. Strictly speaking, the latter
statement has to be restricted to the case where the dimension of ${\mathcal W}_{min}(\hat{\boldsymbol\lambda})$ is
less or equal three, which will be satisfied for all examples considered in this paper.
In the case of one-dimensional ${\mathcal W}_{min}(\hat{\boldsymbol\lambda})$ (collinear ground state) we have
a smooth maximum of  $j_{min}({\boldsymbol\lambda})$, whereas in the cases of a
two- or higher-dimensional ${\mathcal W}_{min}(\hat{\boldsymbol\lambda})$ we have a singular maximum
with a conical structure of $j_{min}({\boldsymbol\lambda})$, at least for some directions in the ${\boldsymbol\lambda}$-space.

According to the above remarks the ground state configuration ${\mathbf s}$ can be written in the form
\begin{equation}\label{T7}
  {\mathbf s}= W\,\Gamma
  \;,
\end{equation}
where $W$ is an $N\times M$-matrix the columns of which span ${\mathcal W}_{min}(\hat{\boldsymbol\lambda})$,
and $\Gamma$ is a real $M\times M$-matrix. For the $N\times N$ {\em Gram matrix}
\begin{equation}\label{T8}
  G\equiv{\mathbf s}\,{\mathbf s}^\top
\end{equation}
we obtain the following representation:
\begin{equation}\label{T9}
 G\stackrel{(\ref{T7},\ref{T8})}{=}\left(  W\,\Gamma\right)\,\left(  W\,\Gamma\right)^\top=W\,\Gamma\,\Gamma^\top\,W^\top\equiv W\,\Delta\,W^\top
\;.
\end{equation}
Here $\Delta=\Gamma\,\Gamma^\top$ is a positive semi-definite real $M\times M$-matrix that can be obtained
as a solution of the inhomogenous system of linear equations
\begin{equation}\label{T10}
  1={\mathbf s}_\mu\cdot{\mathbf s}_\mu=G_{\mu\mu}\stackrel{(\ref{T9})}{=}\left(W\,\Delta\,W^\top\right)_{\mu\mu},\quad \mu=1,\ldots,M
  \;.
\end{equation}
For the examples considered in this paper this system of equations has always a unique solution; for the general case see  \cite{S17a}.

Let $\Gamma=\sqrt{\Delta}\,R$ be the polar decomposition of $\Gamma$ with $R\in O(M)$, then (\ref{T7}) assumes the form
\begin{equation}\label{T11}
  {\mathbf s}=W\,\sqrt{\Delta}\,R
  \;.
\end{equation}
The rotational/reflectional matrix $R$ in (\ref{T11}) can be chosen quite generally
due to the invariance of $ H({\mathbf s})$ under rotations/reflections. If for each pair of
ground states $({\mathbf s},{\mathbf s}')$ there exists an $R\in O(M)$ such that
${\mathbf s}'={\mathbf s}\,R$ then ${\mathbf s}$  will be called {\em essentially unique}.

Next let $\pi\in S_N$ be a permutation and $\Pi$ its standard representation as an $N\times N$-matrix.
Further assume that $\Pi$ is a {\em symmetry}, i.~e., satisfying $\Pi\,{\mathbbm J}={\mathbbm J}\,\Pi$.
Then it can be shown \cite{S17a} that the ground state gauge vector $\hat{\boldsymbol\lambda}$ is
invariant under $\Pi$, i.~e., $\Pi\,\hat{\boldsymbol\lambda}=\hat{\boldsymbol\lambda}$.
Consequently,  $\Pi\,{\mathbbm J}(\hat{\boldsymbol\lambda})={\mathbbm J}(\hat{\boldsymbol\lambda})\,\Pi$.
Moreover, if ${\mathbf s}$ is a ground state then $\Pi\,{\mathbf s}$ will also be a ground state. If
${\mathbf s}$ is essentially unique, as it will be the case in the examples considered in Section \ref{sec:E},
then there exists an $R\in O(M)$ such that $\Pi\,{\mathbf s}={\mathbf s}\,R$, see \cite{S17a}. This means that
the permutation $\pi$ of spin vectors can be compensated by a suitable rotation/ reflection. In this case
${\mathbf s}$ has been called a {\em symmetric ground state} in \cite{SL03}.

In the following we will restrict ourselves to the special case where $\pi$ is a product of disjoint transpositions
such that $\Pi^2={\mathbbm 1}$. Hence the eigenvalues of $\Pi$ (``parities") are $\pm 1$ and the corresponding orthogonal
real eigenspaces $A_\pm$ span ${\mathbbm R}^N$. Moreover, it follows that the two subspaces
$A_\pm$ are invariant under ${\mathbbm J}(\hat{\boldsymbol\lambda})$ and the latter matrix has a block structure
w.~r.~t.~a suitable eigenbasis of $\Pi$. The characteristic polynomial
$p({\boldsymbol\lambda},x)=\det\left( {\mathbbm J}({\boldsymbol\lambda})-x\,{\mathbbm 1}\right)$
will accordingly be split into two factors.

Recall that a ground state ${\mathbf s}$ can be obtained as a linear combination of vectors from
${\mathcal W}_{min}(\hat{\boldsymbol\lambda})$, the eigenspace of ${\mathbbm J}(\hat{\boldsymbol\lambda})$ for the eigenvalue
$\hat{\jmath}$. This results in the following alternative: Either ${\mathcal W}_{min}(\hat{\boldsymbol\lambda})$ is completely
contained in $A_+$ or $A_-$, or ${\mathcal W}_{min}(\hat{\boldsymbol\lambda})$ can be split into two orthogonal subspaces
${\mathcal W}_+(\hat{\boldsymbol\lambda})$ and ${\mathcal W}_-(\hat{\boldsymbol\lambda})$ such that
${\mathcal W}_\pm(\hat{\boldsymbol\lambda})\subset A_\pm$. We will accordingly define:
%%%%%%%%%%%%%%%%%%%%%%%%%%%%%%%%%%%%%%%%%%%%%%%%%%%%%%%%%%%%%%%%%%%%%%%%%%%%%%%%%%%%%%%%%%%%%%%%%%%%%%%%%%%%%%%%%%%%%%%
\begin{defi}\label{D1}
 With the preceding definitions the ground state ${\mathbf s}$  will be called {\em isolated}
 iff ${\mathcal W}_{min}(\hat{\boldsymbol\lambda})\subset A_+$ or ${\mathcal W}_{min}(\hat{\boldsymbol\lambda})\subset A_-$.
 Otherwise, ${\mathbf s}$  will be called {\em cooperative}.
\end{defi}
%%%%%%%%%%%%%%%%%%%%%%%%%%%%%%%%%%%%%%%%%%%%%%%%%%%%%%%%%%%%%%%%%%%%%%%%%%%%%%%%%%%%%%%%%%%%%%%%%%%%%%%%%%%%%%%%%%%%%%%
According to this definition, a collinear ground state is always isolated.
But for the coplanar ground states that will occur in the examples of this paper both possibilities
may be realized as we will show in the next sections.

Before we get into the sections with the examples we would like to make some general remarks about the dimension
of the eigenvalue varieties $j_{min}({\boldsymbol\lambda})$. Let $\Lambda$ denote the $L$-dimensional linear space of
gauge parameters ${\boldsymbol\lambda}$ taking into account (\ref{T4a}) and possible identifications due to symmetries.
In the case of coplanar ground states the eigenvalue
$j_{min}(\hat{\boldsymbol\lambda})$ will be two-fold degenerate and hence it is sensible to consider the subvariety
$j_{min}^{(2)}({\boldsymbol\lambda})$ of points
$({\boldsymbol\lambda},j_{min}({\boldsymbol\lambda})),\,{\boldsymbol\lambda}\in\Lambda$, where $j_{min}({\boldsymbol\lambda})$ is two-fold
degenerate. Its dimension can be estimated by means of certain rules that play a role in quantum mechanics in the context of
avoided level crossing, see \cite{NW29} or \cite{A89}. These rules are obtained by considering the codimension of the manifold $RS_N^{(2)}$
of, in our case, real symmetric $N\times N$-matrices with one pair of two-fold degenerate eigenvalues relative to
the space $RS_N$ of all real symmetric $N\times N$-matrices. This codimension is two, independent of $N$.
(For $N=2$ the space of real symmetric $2\times 2$-matrices is three-dimensional, and the subspace of matrices with degenerate
eigenvalues, necessarily of the form $\left(\begin{array}{cc}
                                        j & 0 \\
                                        0 & j
                                      \end{array}\right)$, is one-dimensional.)

In the ground state problem we have considered the $L$-dimensional subspace of $RS_N$ formed by matrices
${\mathbbm J}({\boldsymbol\lambda}),\; {\boldsymbol\lambda}\in\Lambda$. Hence, {\em in the generic case},
we expect that the sub-manifold of dressed ${\mathbbm J}$-matrices with a pair of two-fold degenerate eigenvalues
will also have the codimension two and hence the dimension $L-2$. The clause ``{\em in the generic case}" means that
in special cases this rule can be violated, see below. It is further plausible that the rule also holds for
the special case of two-fold degenerate minimal eigenvalues and hence the dimension of the variety $j_{min}^{(2)}({\boldsymbol\lambda})$
is also expected to be $L-2$.

In the case of $L=2$ this means that there will be only one point $\hat{\boldsymbol\lambda}$ such that
$j_{min}(\hat{\boldsymbol\lambda})$ will be two-fold degenerate. In the neighbourhood of this point the eigenvalue
variety  $j_{min}({\boldsymbol\lambda})$ will have a conical structure, see, e.~g.,  figure $3$ in \cite{S17a}
for the equilateral spin triangle (without taking into account its symmetry).
A generic curve in the eigenvalue variety  $j_{min}({\boldsymbol\lambda})$ would miss the vertex of the cone
and hence show ``avoided level crossing".
In our example of isolated coplanar ground states of the almost regular cube, see Section \ref{sec:EC},
we will have $L=3$ independent gauge parameters and hence
an $L-2=1$-dimensional variety $j_{min}^{(2)}({\boldsymbol\lambda})$ with a smooth maximum of $j_{min}({\boldsymbol\lambda})$
at $\hat{\boldsymbol\lambda}$.

It is well-known \cite{NW29} that the rules of avoided level crossing will break down in the case of symmetries.
Here we are not {\em in the generic case} since the space $RS_N$ as well as the sub-manifold $RS_N^{(2)}$ would have to
be replaced by sets of matrices commuting with the unitary symmetries. Since our examples in Section \ref{sec:E}
will have an involutary symmetry $\Pi$ we have to review our above arguments for the isolated ground state case.
It suffices to consider the almost regular cube with $N=8$ in Subsection \ref{sec:EC}. Anticipating that the coplanar ground state
is a linear combination of vectors from $A_-$, the eigenspace of $\Pi$ for the eigenvalue $-1$,
we may reformulate the ground state problem for $N/2=4$ independent spin vectors, see Eq.~(\ref{EC5}) and (\ref{EC6}).
If the resulting $\frac{N}{2}\times \frac{N}{2}$-matrix $\tilde{\mathbbm J}$ has no further symmetries we may repeat
the above arguments for the generic dimension $L-2=1$ of the eigenvalue variety $j^{(2)}_{min}({\boldsymbol\lambda})$.

The case of cooperative coplanar ground states is different. Here we have intersections of minimal eigenvalues belonging to different
parities w.~r.~t.~the symmetry $\Pi$ and hence the dimension of $j_{min}^{(2)}({\boldsymbol\lambda})$ may be larger than $L-2$.
This can be illustrated for all three examples below:
For the isosceles triangle in Section \ref{sec:ET} we have $L=1$ and, nevertheless, a coplanar ground state.
For the square with diagonal bond in Section \ref{sec:ES} we have $L=2$ and a one-dimensional variety $j_{min}^{(2)}({\boldsymbol\lambda})$,
see Figure \ref{FIGFG}. Finally, for the almost regular cube in Section \ref{sec:EC} we have $L=3$ and, in the case of cooperative
ground states, a two-dimensional variety $j_{min}^{(2)}({\boldsymbol\lambda})$. In all these cases the cooperative ground states
lie in varieties $j_{min}^{(2)}({\boldsymbol\lambda})$ with codimension one.

%%%%%%%%%%%%%%%%%%%%%%%%%%%%%%%%%%%%%%%%%%%%%%%%%%%%%%%%%%%%%%%%%%%%%%%%%%%%%%%%%%%%%%%%%%%%%%%%%%%%%%%%%%%%%%%%%%%%%%%%%%%%%%%%%%%%%%%%%%
\section{Examples}\label{sec:E}
%%%%%%%%%%%%%%%%%%%%%%%%%%%%%%%%%%%%%%%%%%%%%%%%%%%%%%%%%%%%%%%%%%%%%%%%%%%%%%%%%%%%%%%%%%%%%%%%%%%%%%%%%%%%%%%%%%%%%%%%%%%%%%%%%%%%%%%%%%

%%%%%%%%%%%%%%%%%%%%%%%%%%%%%%%%%%%%%%%%%%%%%%%%%%%%%%%%%%%%%%%%%%%%%%%%%%%%%%%%%%%%%%%%%%%%%%%%%%%%%%%%%%%%%%%%%%%%%%%%%%%%%%%%%%%%%%%%%%
\subsection{Isosceles triangle}\label{sec:ET}
%%%%%%%%%%%%%%%%%%%%%%%%%%%%%%%%%%%%%%%%%%%%%%%%%%%%%%%%%%%%%%%%%%%%%%%%%%%%%%%%%%%%%%%%%%%%%%%%%%%%%%%%%%%%%%%%%%%%%%%%%%%%%%%%%%%%%%%%%%
\begin{figure}[ht]
  \centering
    \includegraphics[width=1.0\linewidth]{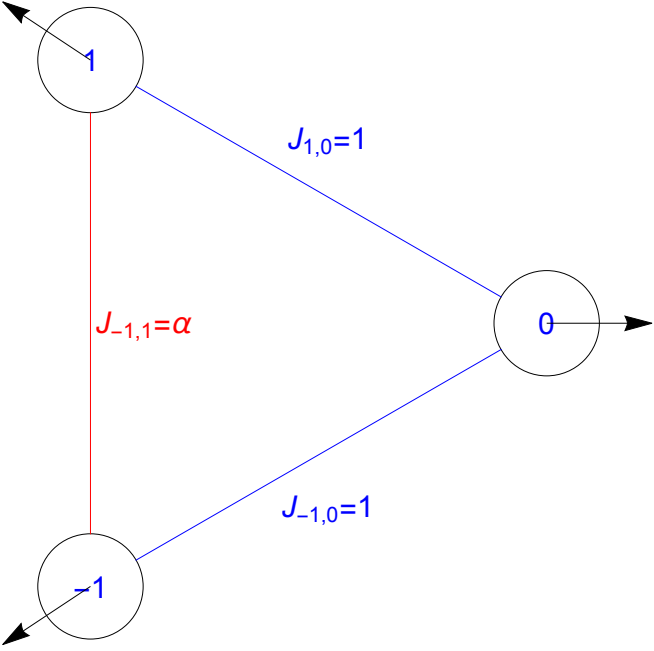}
  \caption[Example S]
  {The spin triangle with $J_{1,0}=J_{-1,0}=1$ and a variable bond $J_{-1,1}=\alpha$.
  The arrows indicate the coplanar ground state for $\alpha=0.6$, see (\ref{ET10}).
  }
  \label{FIGTRI}
\end{figure}

As one of the simplest examples illustrating the considerations of Section \ref{sec:T}
we consider three spins ${\mathbf s}_\mu$ indexed by $\mu\in\{-1,0,1\}$, two AF couplings $J_{1,0}=J_{-1,0}=1$ and a variable bond $J_{-1,1}=\alpha$,
see Figure \ref{FIGTRI}.
% The unusual numeration of the spins is adapted to the more general case of an almost regular odd spin ring
%in Section \ref{sec:ER}.
Due to the symmetry w.~r.~t.~the transposition $(1\leftrightarrow -1)$
and (\ref{T4a})
its dressed ${\mathbbm J}$-matrix assumes the form
\begin{equation}\label{ET1}
 {\mathbbm J}(\lambda)=\left(
\begin{array}{ccc}
 \lambda  & 1 & \alpha  \\
 1 & -2 \lambda  & 1 \\
 \alpha  & 1 & \lambda  \\
\end{array}
\right)\;.
\end{equation}
For negative $\alpha$ this system admits an essentially unique collinear ground state symbolically written
as ${\mathbf s}_{coll}=(\uparrow\downarrow\uparrow)$. If the variable bond  has a small positive value, ${\mathbf s}_{coll}$
remains the ground state.

The eigenvalues $j_\mu(\lambda)$ and eigenvectors ${\mathbf e}_\mu(\lambda)$ of (\ref{ET1}) can be analytically determined:
\begin{eqnarray}
\label{ET2a}
  j_{-1}(\lambda) &=& \lambda -\alpha\\
\label{ET2b}
 {\mathbf e}_{-1}(\lambda)&=&(-1,0,1)^\top\\
\label{ET2c}
  j_{0}(\lambda) &=& \frac{1}{2} \left(\alpha -\lambda -\sqrt{(\alpha +3 \lambda )^2+8}\right)\\
\label{ET2d}
  {\mathbf e}_{0}(\lambda)&=&
\left(1,\frac{1}{2} \left(-\alpha -3 \lambda -\sqrt{(\alpha +3 \lambda    )^2+8}\right),1\right)^\top\\
\label{ET2e}
 j_{1}(\lambda) &=& \frac{1}{2} \left(\alpha -\lambda +\sqrt{(\alpha +3 \lambda )^2+8}\right)\\
\label{ET2f}
 {\mathbf e}_{1}(\lambda)&=&\left(1,\frac{1}{2} \left(-\alpha -3 \lambda +\sqrt{(\alpha +3\lambda )^2+8}\right),1\right)^\top
 \;.
\end{eqnarray}

The second eigenvalue $j_0(\lambda)$ has a smooth maximum at
\begin{equation}\label{ET3}
  \lambda_0=-\frac{1}{3} (\alpha +1)
\;,
\end{equation}
and intersects the line $j_{-1}(\lambda)$ at
\begin{equation}\label{ET4}
  \lambda_1= \frac{\alpha ^2-1}{3 \alpha }
\;.
\end{equation}
The equation $\lambda_0= \lambda_1$ defines the critical value
\begin{equation}\label{ET5}
  \alpha_c=\frac{1}{2}
\;.
\end{equation}

\begin{figure}[ht]
  \centering
    \includegraphics[width=0.7\linewidth]{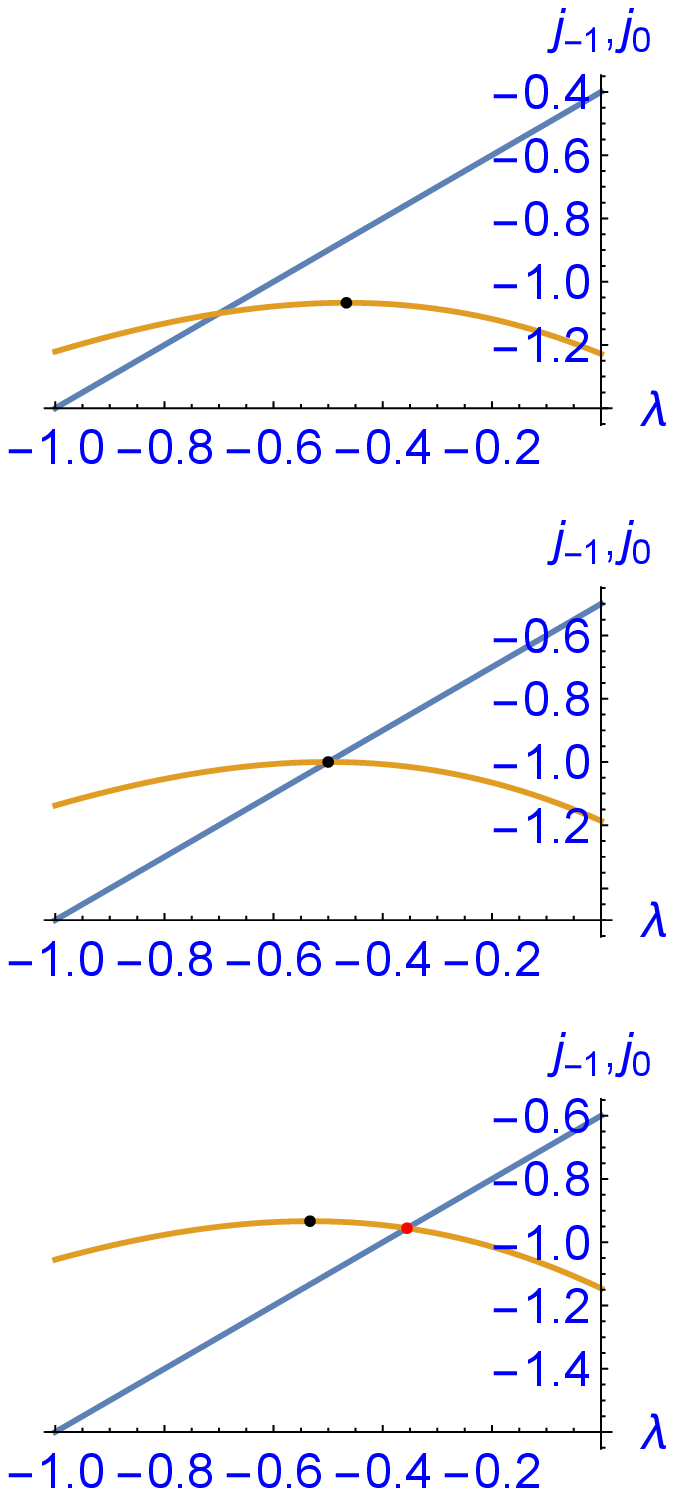}
  \caption[Example S]
  {The intersection between $j_{-1}(\lambda)$ (blue line)
  and $j_{0}(\lambda)$ (orange curve) for the values $\alpha=0.4$, $\alpha=0.5$ and $\alpha=0.6$
  (from top to bottom). The smooth maximum of $j_0( \lambda)$ is indicated by a black dot, the maximum of
  $j_{min}(\lambda)$ by a red dot if it differs from that of $j_{-1}(\lambda)$.
  }
  \label{FIGFG1}
\end{figure}

It turns out that $j_{min}(\lambda)=\mbox{Min}\left(j_{-1}(\lambda),j_0(\lambda)\right)$.
If $\lambda_0>\lambda_1$ or, equivalently, $\alpha<\alpha_c=1/2$, then the smooth maximum of $j_{-1}(\lambda)$
coincides with the maximum of  $j_{min}(\lambda)$, see Figure \ref{FIGFG1}, top panel.
If $\lambda_0<\lambda_1$ or, equivalently, $\alpha>\alpha_c=1/2$, then the maximum of  $j_{min}(\lambda)$
occurs at the intersection of $j_{-1}(\lambda)$ and $j_0(\lambda)$ , see Figure \ref{FIGFG1}, bottom panel.
The boundary between these cases is given by the equation $\lambda_0= \lambda_1$ or, equivalently, $\alpha=\alpha_c=1/2$
and corresponds to the case where the smooth maximum of $j_{-1}(\lambda)$ coincides with the
intersection of $j_{-1}(\lambda)$ and $j_0(\lambda)$ , see Figure \ref{FIGFG1}, middle panel.

\begin{figure}[t]
  \centering
    \includegraphics[width=1.0\linewidth]{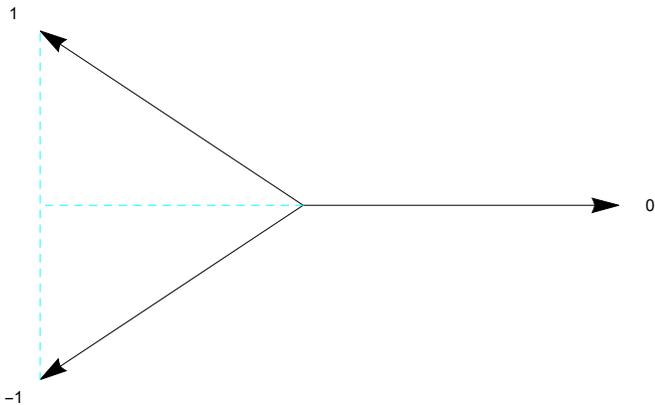}
  \caption
  {Coplanar ground state of the isosceles triangle with  bond $\alpha=0.6$, see also Figure \ref{FIGTRI}.
    We observe that ${\mathbf s}_1=R\,{\mathbf s}_{-1}$ where $R$ is the reflection
   on the axis spanned by ${\mathbf s}_0$.
  }
  \label{FIGGS1}
\end{figure}

It is clear that in the case $\alpha>\alpha_c=1/2$ the eigenspace ${\mathcal W}_{min}(\hat{\lambda})$ of
${\mathbbm J}(\hat{\lambda})$ is two-dimensional and hence we expect a coplanar ground state in this case.
To verify this expectation we consider the matrix
\begin{equation}\label{ET6}
  W=\left(
\begin{array}{cc}
 -1 & -1 \\
 0 & 2 \alpha  \\
 1 & -1 \\
\end{array}
\right)
\;,
\end{equation}
the columns of which span ${\mathcal W}_{min}(\hat{\lambda})$ such that the first column is an eigenvector
of $\Pi$ with eigenvalue $-1$ and the second one analogously with eigenvalue $+1$.
According to Section \ref{sec:T} we have to solve the system of equations (\ref{T10}) for the
unknown matrix entries of
\begin{equation}\label{ET7}
  \Delta =\left(
\begin{array}{cc}
 \delta _{11} & \delta _{12} \\
 \delta _{12} & \delta _{22} \\
\end{array}
\right).
\end{equation}
The unique solution reads
\begin{equation}\label{ET8}
 \Delta =\left(
\begin{array}{cc}
 -\frac{1-4 \alpha ^2}{4 \alpha ^2} & 0 \\
 0 & \frac{1}{4 \alpha ^2} \\
\end{array}
\right)
\end{equation}
and yields a ground state Gram matrix of the form
\begin{equation}\label{ET9}
 G=\left(
\begin{array}{ccc}
 1 & -\frac{1}{2 \alpha } & \frac{1}{2 \alpha ^2}-1 \\
 -\frac{1}{2 \alpha } & 1 & -\frac{1}{2 \alpha } \\
 \frac{1}{2 \alpha ^2}-1 & -\frac{1}{2 \alpha } & 1 \\
\end{array}
\right).
\end{equation}
A possible ground state configuration compatible with this Gram matrix can be, according to (\ref{T11}),
chosen as
\begin{equation}\label{ET10}
{\mathbf s}=W\,\sqrt{\Delta}\,R=\left(
\begin{array}{cc}
 -\frac{1}{2 \alpha } & -\frac{1}{2} \sqrt{4-\frac{1}{\alpha ^2}} \\
 1 & 0 \\
 -\frac{1}{2 \alpha } & \frac{1}{2} \sqrt{4-\frac{1}{\alpha ^2}} \\
\end{array}
\right),
\end{equation}
where $R\in SO(2)$ represents a rotation with the angle $\pi/2$.
An example for $\alpha=0.6$ is shown in Figure \ref{FIGGS1}.

We will investigate the isolated/cooperative nature of the ground states. The symmetry in question is the
linear representation $\Pi$ of the transposition $\pi=(1\leftrightarrow -1)$.
According to (\ref{ET2b}) the eigenvector ${\mathbf e}_{-1}(\lambda)=(-1,0,1)^\top$  is independent of
${\lambda}$ and has the eigenvalue $-1$ w.~r.~t.~$\Pi$. The other two eigenvectors of
${\mathbbm J}(\hat{\lambda})$ belong to the eigenspace of $\Pi$ corresponding to the eigenvalue $+1$.
For $\alpha>\alpha_c=1/2$ the ground state ${\mathbf s}$ is a proper superposition of the first two eigenvectors
and hence these ground states are cooperative according to Definition \ref{D1}.

For  $\alpha>\alpha_c=1/2$ all coplanar ground states have the same symmetry properties as the
ground state shown in Figure \ref{FIGGS1} for $\alpha =0.6$:
The spin vector ${\mathbf s}_1$ is given by the reflection $R\,{\mathbf s}_{-1}$ on the axis spanned by ${\mathbf s}_0$.

The reasons for these symmetry properties can best be understood by exploiting the fact that ${\mathbf s}$
is a symmetric ground state w.~r.~t.~$\Pi$. This means that $\Pi\,{\mathbf s}={\mathbf s}\,R$
with a suitable $R\in O(2)$, see Section \ref{sec:T}. ${\mathbf s}_0$ is left fixed
by $\Pi$, hence $R={\mathbbm 1}$ or $R$ is the reflection on the axis spanned by ${\mathbf s}_0$.
The first possibility would mean that ${\mathbf s}$ is isolated contrary to what we have already established.
We thus conclude that the swapping of ${\mathbf s}_1$ and ${\mathbf s}_{-1}$ is compensated by
the reflection $R$ as shown in Figure \ref{FIGGS1}.

\begin{figure}[ht]
  \centering
    \includegraphics[width=1.0\linewidth]{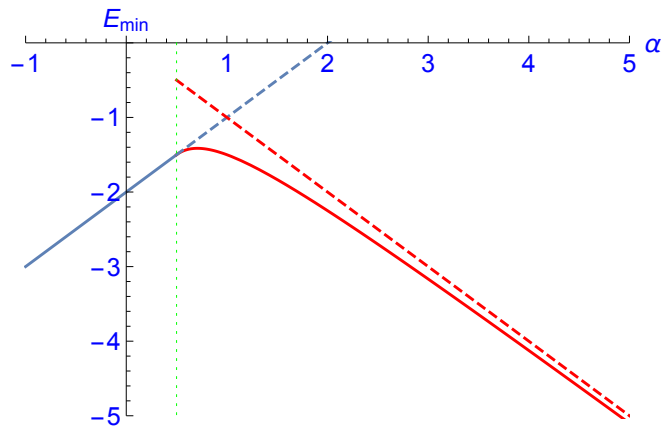}
  \caption
  {The ground state energy $E_{min}$  as a function of the diagonal bond coupling parameter $\alpha$.
  For $\alpha\le\alpha_c=1/2$ the minimal energy is given by (\ref{ET11}) (blue line) according to the collinear ground state ${\mathbf s}_{coll}$.
  For $\alpha >\alpha_c=1/2$ the coplanar cooperative ground state with minimal energy (\ref{ET12}) assumes the role of the ground state (red curve).
  Its asymptotic limit (\ref{ET13}) is given by the red dashed line.
  }
  \label{FIGEN1}
\end{figure}

Finally, we consider the ground state energy as a function of $\alpha$. For $\alpha\le \alpha_c=1/2$
the collinear ground state ${\mathbf s}_{coll}=(\uparrow\downarrow\uparrow)$ has
the energy
\begin{equation}\label{ET11}
  E_{min}=\alpha-2
\;,
\end{equation}
see the blue line in Figure \ref{FIGEN1}.
For $\alpha> \alpha_c=1/2$ the coplanar ground state energy can be calculated as
\begin{equation}\label{ET12}
   E_{min}=-\alpha -\frac{1}{2 \alpha }
\;,
\end{equation}
see the red curve in Figure \ref{FIGEN1}.
In the asymptotic limit $\alpha\rightarrow\infty$ the ground state approaches the form
${\mathbf s}=((1,0),(0,1),(-1,0))^\top$  with a minimal energy of
\begin{equation}\label{ET13}
   E_{min}=-\alpha
\;,
\end{equation}
see the red dashed line in Figure \ref{FIGEN1}.

%%%%%%%%%%%%%%%%%%%%%%%%%%%%%%%%%%%%%%%%%%%%%%%%%%%%%%%%%%%%%%%%%%%%%%%%%%%%%%%%%%%%%%%%%%%%%%%%%%%%%%%%%%%%%%%%%%%%%%%%%%%%%%%%%%%%%%%%%%
\subsection{Square with diagonal bond}\label{sec:ES}
%%%%%%%%%%%%%%%%%%%%%%%%%%%%%%%%%%%%%%%%%%%%%%%%%%%%%%%%%%%%%%%%%%%%%%%%%%%%%%%%%%%%%%%%%%%%%%%%%%%%%%%%%%%%%%%%%%%%%%%%%%%%%%%%%%%%%%%%%%

\begin{figure}[ht]
  \centering
    \includegraphics[width=1.0\linewidth]{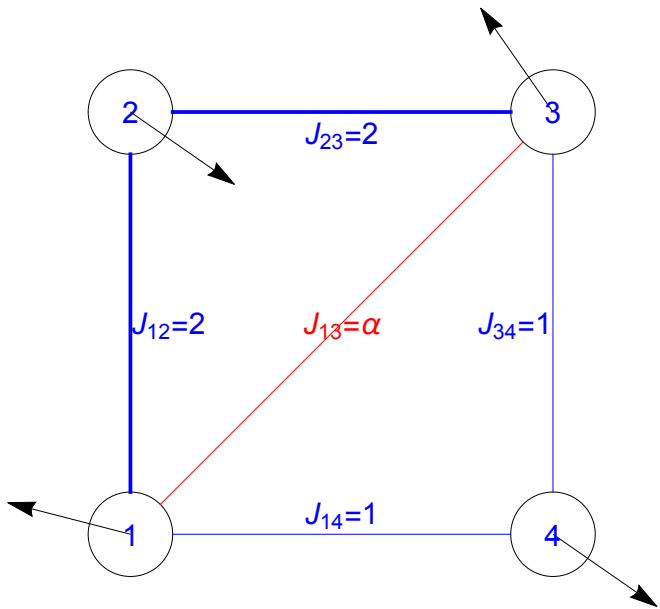}
  \caption[Example S]
  {The spin square with $J_{12}=J_{23}=2$ and  $J_{34}=J_{41}=1$ and a variable diagonal bond $J_{13}=\alpha$.
  The arrows indicate the coplanar ground state for $\alpha=1.6$, see (\ref{ES10}).
  }
  \label{FIGSQUARE}
\end{figure}
We consider a square with AF couplings $J_{12}=J_{23}=2$ and  $J_{34}=J_{41}=1$ and a variable diagonal bond $J_{13}=\alpha$,
see Figure \ref{FIGSQUARE}.
Due to the symmetry w.~r.~t.~the transposition $(1\leftrightarrow 3)$
and (\ref{T4a})
its dressed ${\mathbbm J}$-matrix assumes the form
\begin{equation}\label{ES1}
 {\mathbbm J}(\boldsymbol \lambda)=\left(
\begin{array}{cccc}
 \lambda _1 & 2 & \alpha  & 1 \\
 2 & \lambda _2 & 2 & 0 \\
 \alpha  & 2 & \lambda _1 & 1 \\
 1 & 0 & 1 & -2 \lambda _1-\lambda _2 \\
\end{array}
\right)\;.
\end{equation}
The AF square without diagonal bond is a bipartite system and admits an essentially unique collinear ground state symbolically written
as ${\mathbf s}_{coll}=(\uparrow\downarrow\uparrow\downarrow)$. If a negative diagonal bond $\alpha$ is added to the square, ${\mathbf s}_{coll}$
remains the ground state. This holds even for small positive values of $\alpha$.

\begin{figure}[ht]
  \centering
    \includegraphics[width=1.0\linewidth]{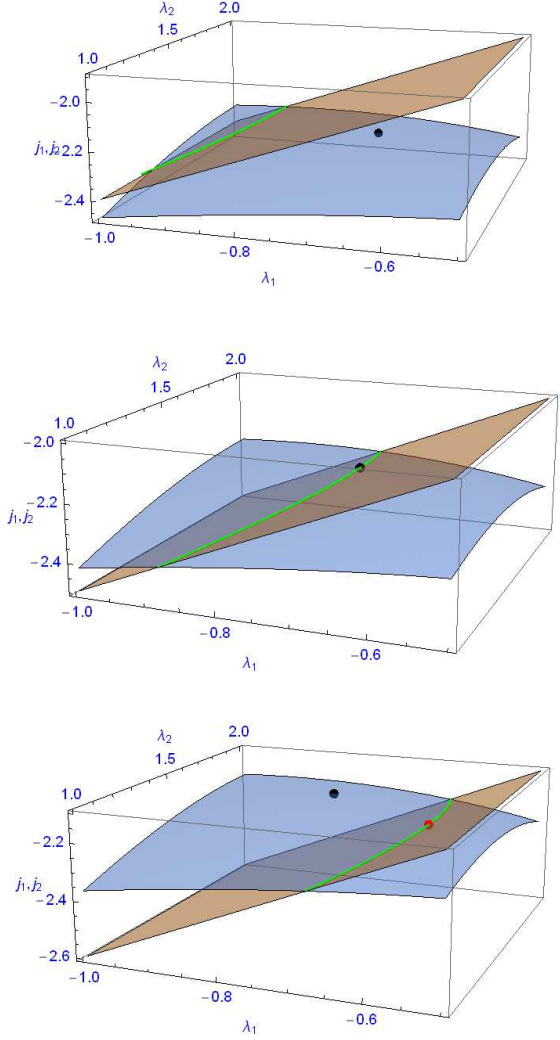}
  \caption[Example S]
  {The intersection between $j_1(\boldsymbol \lambda)$ (dark orange plane)
  and $j_2(\boldsymbol \lambda)$ (blue surface) for the values $\alpha=1.4$, $\alpha=1.5$ and $\alpha=1.6$
  (from top to bottom) marked by a green curve. The smooth maximum of $j_2(\boldsymbol \lambda)$ is indicated by a black dot, the maximum of
  $j_{min}(\boldsymbol \lambda)$ by a red dot if it differs from that of $j_2(\boldsymbol \lambda)$.
  }
  \label{FIGFG}
\end{figure}

The eigenvalues of (\ref{ES1}) can be analytically determined.
The first one reads $j_1(\boldsymbol \lambda)=\lambda_1-\alpha$ with eigenvector $(-1, 0, 1, 0)^\top$, but the other
three eigenvalues are too involved to be represented here. The collinear ground state corresponds to a smooth maximum
of a certain eigenvalue henceforward denoted by $j_2(\boldsymbol \lambda)$. This maximum is attained at
\begin{equation}\label{ES2}
 \lambda_1=-\frac{\alpha}{2},\; \lambda_2=1+ \frac{\alpha}{2}
 \;,
\end{equation}
 see Figure \ref{FIGFG}. It turns out that the minimal eigenvalue $j_{min}(\boldsymbol \lambda)$
 is given by $\mbox{Min}\left(j_1(\boldsymbol \lambda),j_2(\boldsymbol \lambda)\right)$.
 Recall that the ground states can be obtained by the unique maximum of $j_{min}(\boldsymbol \lambda)$.
 If the smooth maximum of $j_2(\boldsymbol \lambda)$ is also the maximum of $j_{min}(\boldsymbol \lambda)$
 the ground state is collinear, see the top panel of Figure \ref{FIGFG}. Otherwise the
 maximum of $j_{min}(\boldsymbol \lambda)$ can be found at the highest point of the curve given by
 \onecolumngrid
 \begin{equation}\label{ES3}
 \lambda_1=-\frac{\sqrt{4 \alpha  \left(\alpha +2 \lambda _2\right) \left(\alpha ^2+2 \alpha
   \lambda _2-11\right)+169}-2 \alpha  \left(2 \alpha +\lambda _2\right)+13}{6    \alpha },
 \end{equation}
\twocolumngrid
\noindent that represents the intersection between $j_1(\boldsymbol \lambda)$ and $j_2(\boldsymbol \lambda)$,
see the bottom  panel of Figure \ref{FIGFG}. This case corresponds to a coplanar ground state as we will see below.
The transition between the two cases is given by the condition that the smooth maximum of $j_2(\boldsymbol \lambda)$
lies on the curve (\ref{ES3}), see the middle panel of Figure  \ref{FIGFG}. This condition gives the critical value
$\alpha_c>0$ as the positive common solution of (\ref{ES2}) and (\ref{ES3}):
\begin{equation}\label{ES4}
  \alpha_c=\frac{3}{2}
  \;.
\end{equation}

\begin{figure}[t]
  \centering
    \includegraphics[width=1.0\linewidth]{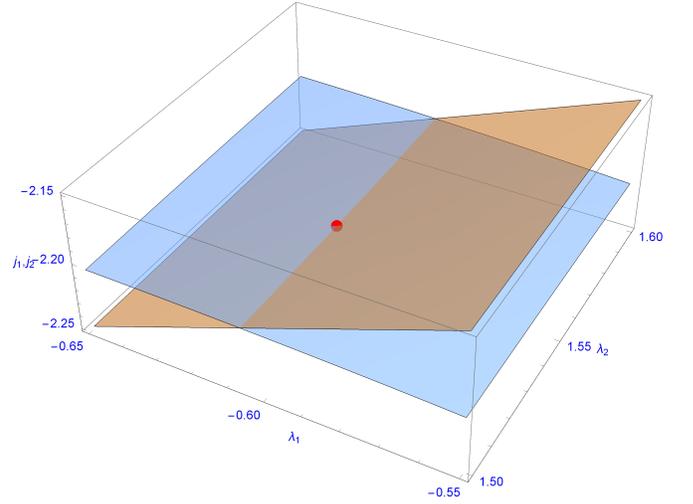}
  \caption[Example S]
  {The infinitesimal neighbourhood of the maximum of $j_{min}(\boldsymbol \lambda)$ for $\alpha=1.6$
  corresponding to the bottom panel of Figure \ref{FIGFG}.
  Contrary to appearances the blue plane is not horizontal but has a small slope.
  }
  \label{FIGEDGE}
\end{figure}

The case $\alpha>\alpha_c$ is remarkable in that it differs from the usual picture where the graph of $j_{min}(\boldsymbol \lambda)$
has a conical structure in the infinitesimal neighbourhood of its maximum which leads to the above mentioned effect called
``avoided level crossing".
In our case the tangent structure of $j_{min}(\boldsymbol \lambda)$ at its maximum is rather a wedge with
a horizontal edge, see Figure \ref{FIGEDGE}.

\begin{figure}[ht]
  \centering
    \includegraphics[width=1.0\linewidth]{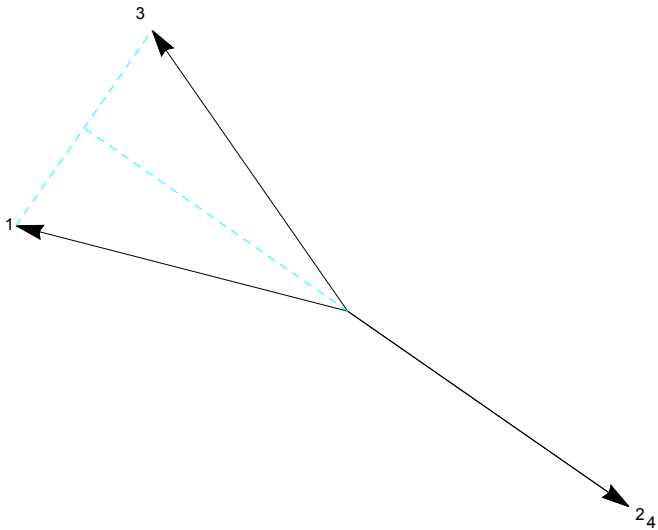}
  \caption
  {Coplanar ground state of the square with diagonal bond $\alpha=1.6$, see also Figure \ref{FIGSQUARE}.
    We observe that ${\mathbf s}_2={\mathbf s}_4$ and that ${\mathbf s}_3=R\,{\mathbf s}_1$ where $R$ is the reflection
   on the axis spanned by ${\mathbf s}_2={\mathbf s}_4$.
  }
  \label{FIGGS}
\end{figure}

It remains to determine the coplanar ground states in the case $\alpha>\alpha_c=\frac{3}{2}$.
After a short calculation it follows that at the curve given by
(\ref{ES3}) the eigenvalue $j_1(\boldsymbol \lambda)=j_2(\boldsymbol \lambda)$
assumes its maximum at
\begin{equation}\label{ES5}
 \hat{\lambda}_2=\frac{15-2 \alpha ^2}{4 \alpha }
 \;.
\end{equation}
The corresponding two-dimensional eigenspace ${\mathcal W}_{min}(\hat{\boldsymbol\lambda})$ of ${\mathbbm J}(\hat{\boldsymbol\lambda})$ is spanned by the columns of
\begin{equation}\label{ES6}
  W=\left(
\begin{array}{cc}
 -\frac{3}{\alpha } & -1 \\
 1 & 0 \\
 0 & 1 \\
 1 & 0 \\
\end{array}
\right).
\end{equation}
According to Section \ref{sec:T} we have to solve the system of equations (\ref{T10}) for the
unknown matrix entries of
\begin{equation}\label{ES7}
  \Delta =\left(
\begin{array}{cc}
 \delta _{11} & \delta _{12} \\
 \delta _{12} & \delta _{22} \\
\end{array}
\right).
\end{equation}
The unique solution reads
\begin{equation}\label{ES8}
 \Delta =\left(
\begin{array}{cc}
 1 & -\frac{3}{2 \alpha } \\
 -\frac{3}{2 \alpha } & 1 \\
\end{array}
\right)
\end{equation}
and yields a ground state Gram matrix of the form
\begin{equation}\label{ES9}
 G=\left(
\begin{array}{cccc}
 1 & -\frac{3}{2 \alpha } & \frac{9}{2 \alpha ^2}-1 & -\frac{3}{2 \alpha } \\
 -\frac{3}{2 \alpha } & 1 & -\frac{3}{2 \alpha } & 1 \\
 \frac{9}{2 \alpha ^2}-1 & -\frac{3}{2 \alpha } & 1 & -\frac{3}{2 \alpha } \\
 -\frac{3}{2 \alpha } & 1 & -\frac{3}{2 \alpha } & 1 \\
\end{array}
\right).
\end{equation}
A possible ground state configuration compatible with this Gram matrix can, according to (\ref{T11}),
be chosen as
\begin{equation}\label{ES10}
{\mathbf s}=W\,\sqrt{\Delta}=\left(
\begin{array}{cc}
 \frac{a (\alpha -3)-b (\alpha +3)}{2 \sqrt{2} \alpha } & -\frac{a (\alpha -3)+b
   (\alpha +3)}{2 \sqrt{2} \alpha } \\
 \frac{a+b}{2 \sqrt{2}} & \frac{b-a}{2 \sqrt{2}} \\
 \frac{b-a}{2 \sqrt{2}} & \frac{a+b}{2 \sqrt{2}} \\
 \frac{a+b}{2 \sqrt{2}} & \frac{b-a}{2 \sqrt{2}} \\
\end{array}
\right),
\end{equation}
where
\begin{equation}\label{ES11}
  a\equiv \sqrt{2+\frac{3}{\alpha}},\quad \mbox{and } b\equiv \sqrt{2-\frac{3}{\alpha}}
  \;.
\end{equation}
An example for $\alpha=1.6$ is shown in Figure \ref{FIGGS}.

We will investigate the isolated/cooperative nature of the ground states. The symmetry in question is the
linear representation $\Pi$ of the transposition $\pi=(1\leftrightarrow 3)$.
We have already mentioned the eigenvector $(-1,0,1,0)^\top$ of $j_1({\boldsymbol\lambda})$ that is independent of
${\boldsymbol\lambda}$ and has the eigenvalue $-1$ w.~r.~t.~$\Pi$. The other three eigenvectors of
${\mathbbm J}(\hat{\boldsymbol\lambda})$ belong to the eigenspace of $\Pi$ corresponding to the eigenvalue $+1$.
For $\alpha>\alpha_c=3/2$ the ground state ${\mathbf s}$ is a proper superposition of $e_{-1}(\lambda)$ and $e_{0}(\lambda)$
and hence these ground states are cooperative according to Definition \ref{D1}.

For  $\alpha>\alpha_c=3/2$ all coplanar ground states have the same symmetry properties as the
ground state shown in Figure \ref{FIGGS} for $\alpha =1.6$: Two spin vectors coincide, ${\mathbf s}_2={\mathbf s}_4$,
and ${\mathbf s}_3$ is given by the reflection $R\,{\mathbf s}_1$ on the axis given by ${\mathbf s}_2={\mathbf s}_4$.

The reasons for these symmetry properties can best be understood by exploiting the fact that ${\mathbf s}$
is a symmetric ground state w.~r.~t.~$\Pi$. This means that $\Pi\,{\mathbf s}={\mathbf s}\,R$
with a suitable $R\in O(2)$, see Section \ref{sec:T}. ${\mathbf s}_2$ and ${\mathbf s}_4$ are left fixed
by $\Pi$, hence $R={\mathbbm 1}$ if $({\mathbf s}_2,{\mathbf s}_4)$ would be linearly independent.
This would mean that ${\mathbf s}$ is isolated contrary to what we have already established.
We thus conclude ${\mathbf s}_2={\mathbf s}_4$ since ${\mathbf s}_2=-{\mathbf s}_4$ can be excluded by
arguing with the requirement of minimal energy. Further, we conclude that $R$ leaves
${\mathbf s}_2$ invariant and hence must be a reflection on the line given by multiples of ${\mathbf s}_2$.

\begin{figure}[t]
  \centering
    \includegraphics[width=1.0\linewidth]{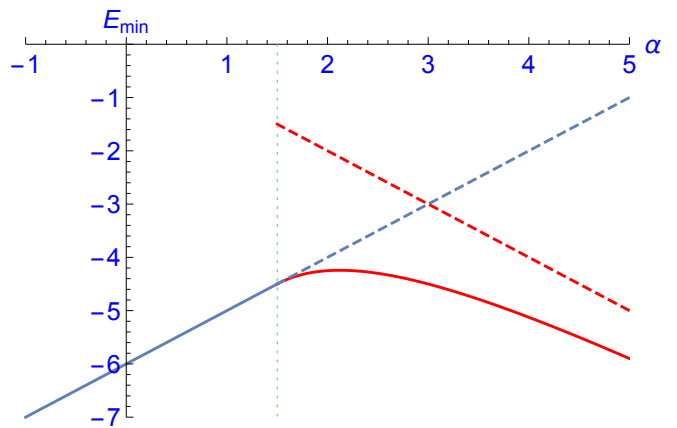}
  \caption
  {The ground state energy $E_{min}$  as a function of the diagonal bond coupling parameter $\alpha$.
    The critical value  $\alpha_c=3/2$ is indicated by a vertical dotted green line.
   For $\alpha\le\alpha_c$ the minimal energy is given by (\ref{ES12}) (blue line) according to the collinear ground state ${\mathbf s}_{coll}$.
  For $\alpha >\alpha_c$ the coplanar cooperative ground state with minimal energy (\ref{ES13}) assumes the role of the ground state (red curve).
  Its asymptotic limit (\ref{ES14}) is given by the red dashed line.
  }
  \label{FIGEN}
\end{figure}

Finally, we consider the ground state energy as a function of $\alpha$. For $\alpha\le \alpha_c=3/2$ the collinear ground state
${\mathbf s}_{coll}=(\uparrow\downarrow\uparrow\downarrow)$  has
the energy
\begin{equation}\label{ES12}
  E_{min}=\alpha-6
\;,
\end{equation}
see the blue line in Figure \ref{FIGEN}.
For $\alpha> \alpha_c=3/2$ the coplanar ground state energy can be calculated as
\begin{equation}\label{ES13}
   E_{min}=-\alpha -\frac{9}{2 \alpha }
\;,
\end{equation}
see the red curve in Figure \ref{FIGEN}.
In the asymptotic limit $\alpha\rightarrow\infty$ the ground state approaches the form
${\mathbf s}=((1,0),(0,1),(-1,0),(0,1))^\top$  with a minimal energy of
\begin{equation}\label{ES14}
   E_{min}=-\alpha
\;,
\end{equation}
see the red dashed line in Figure \ref{FIGEN}.

%%%%%%%%%%%%%%%%%%%%%%%%%%%%%%%%%%%%%%%%%%%%%%%%%%%%%%%%%%%%%%%%%%%%%%%%%%%%%%%%%%%%%%%%%%%%%%%%%%%%%%%%%%%%%%%%%%%%%%%%%%%%%%%%%%%%%%%%%%
\subsection{Almost regular Cube}\label{sec:EC}
%%%%%%%%%%%%%%%%%%%%%%%%%%%%%%%%%%%%%%%%%%%%%%%%%%%%%%%%%%%%%%%%%%%%%%%%%%%%%%%%%%%%%%%%%%%%%%%%%%%%%%%%%%%%%%%%%%%%%%%%%%%%%%%%%%%%%%%%%%

\begin{figure}[t]
  \centering
    \includegraphics[width=1.0\linewidth]{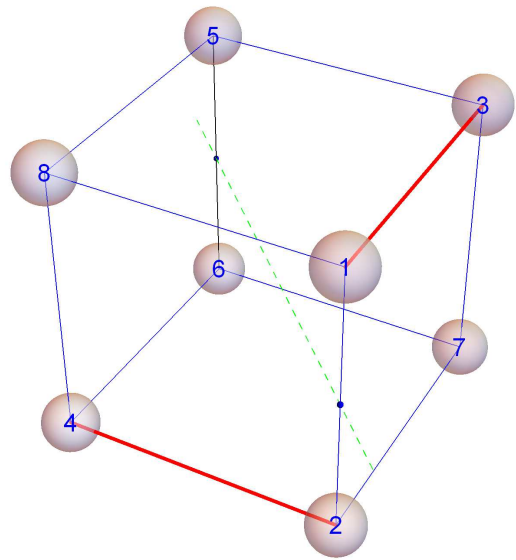}
  \caption
  {The cube with AF coupling $J_{\mu\nu}=1$ (blue lines) except one variable coupling parameter $J_{24}=J_{13}=\alpha$ (red thick lines).
  This spin system is invariant under the rotation about the dashed green axis with an angle of $180^\circ$.
  }
  \label{FIGCUBE}
\end{figure}

We consider a cube with AF coupling  $J_{\mu\nu}=1$  except one variable coupling parameter $J_{24}=J_{13}=\alpha$, see
Figure \ref{FIGCUBE}. The system has the symmetry $\pi=(1\leftrightarrow 2,3\leftrightarrow 4,5\leftrightarrow 6,7\leftrightarrow 8)$
that can be geometrically interpreted as a rotation about a suitable axis with an angle of $180^\circ$, see Figure \ref{FIGCUBE}.
As usual, $\Pi$ denotes the standard representation of $\pi$ by an $8\times 8$-matrix.
The enumeration of the spin sites is adapted to this symmetry.
The AF cube ($\alpha>0$) admits a collinear ground state ${\mathbf s}_{coll}$ symbolically written as
$(\uparrow\downarrow\downarrow\uparrow\uparrow\downarrow\uparrow\downarrow)$ that remains the ground state even for small negative values of $\alpha$.

Taking into account the symmetry $\pi$
the dressed ${\mathbbm J}$-matrix assumes the form
\begin{equation}\label{EC1}
{\mathbbm J}({\boldsymbol\lambda})=
\left(
\begin{array}{cccccccc}
 \lambda _1 & 1 & \alpha  & 0 & 0 & 0 & 0 & 1 \\
 1 & \lambda _1 & 0 & \alpha  & 0 & 0 & 1 & 0 \\
 \alpha  & 0 & \lambda _2 & 0 & 1 & 0 & 1 & 0 \\
 0 & \alpha  & 0 & \lambda _2 & 0 & 1 & 0 & 1 \\
 0 & 0 & 1 & 0 & \lambda _3 & 1 & 0 & 1 \\
 0 & 0 & 0 & 1 & 1 & \lambda _3 & 1 & 0 \\
 0 & 1 & 1 & 0 & 0 & 1 &\lambda_4 & 0 \\
 1 & 0 & 0 & 1 & 1 & 0 & 0 & \lambda_4 \\
\end{array}
\right)\;,
\end{equation}
where $\lambda_4\equiv -\lambda _1-\lambda _2-\lambda _3$ due to (\ref{T4a}).
If we transform ${\mathbbm J}({\boldsymbol\lambda})$ to the eigenbasis of $\Pi$ that is formed by the
columns of the following matrix
\begin{equation}\label{EC1a}
 U=\frac{1}{\sqrt{2}}
\left(
\begin{array}{rrrrrrrr}
 1 & 0 & 0 & 0 & 1 & 0 & 0 & 0 \\
 1 & 0 & 0 & 0 & -1 & 0 & 0 & 0 \\
 0 & 1 & 0 & 0 & 0 & 1 & 0 & 0 \\
 0 & 1 & 0 & 0 & 0 & -1 & 0 & 0 \\
 0 & 0 & 1 & 0 & 0 & 0 & 1 & 0 \\
 0 & 0 & 1 & 0 & 0 & 0 & -1 & 0 \\
 0 & 0 & 0 & 1 & 0 & 0 & 0 & 1 \\
 0 & 0 & 0 & 1 & 0 & 0 & 0 & -1 \\
\end{array}
\right)
\;,
\end{equation}
we obtain the block matrix
\begin{equation}\nonumber
U^\top\,{\mathbbm J}({\boldsymbol\lambda})\,U=
\end{equation}
\begin{equation}\label{EC1}
\left(
\begin{array}{cccccccc}
 \lambda _1+1 & \alpha  & 0 & 1 & 0 & 0 & 0 & 0 \\
 \alpha  & \lambda _2 & 1 & 1 & 0 & 0 & 0 & 0 \\
 0 & 1 & \lambda _3+1 & 1 & 0 & 0 & 0 & 0 \\
 1 & 1 & 1 & \lambda _4 & 0 & 0 & 0 & 0 \\
 0 & 0 & 0 & 0 & \lambda _1-1 & \alpha  & 0 & -1 \\
 0 & 0 & 0 & 0 & \alpha  & \lambda _2 & 1 & 1 \\
 0 & 0 & 0 & 0 & 0 & 1 & \lambda _3-1 & -1 \\
 0 & 0 & 0 & 0 & -1 & 1 & -1 & \lambda _4 \\
\end{array}
\right).
\end{equation}

Its characteristic polynomial $p(x,{\boldsymbol\lambda})$
can hence be split into two factors $p= p_1\,p_2$, such that the zeroes of $p_1$ are the eigenvalues
of parity $+1$ and the zeroes of $p_2$ those of parity $-1$.
The data for the collinear ground state
can be directly obtained by inserting ${\mathbf s}_{coll}$
into (\ref{T3}) and calculating the $\lambda_i$ and the minimal eigenvalue of ${\mathbbm J}({\boldsymbol\lambda})$.
The result reads
\begin{eqnarray}\label{EC2a}
\hat{\jmath}&=& \frac{1}{2} (-\alpha -5),\quad
E_{min}=2 (-\alpha-5),\\
\label{EC2b}
\lambda_1&=&\lambda_2=\frac{\alpha-1}{2},\quad
\lambda_3=\lambda_4=\frac{1-\alpha}{2}\;.
\end{eqnarray}
Inserting these values of the components of ${\boldsymbol\lambda}$ into $p_2$ gives a polynomial $p_2(x,\alpha)$.
The condition $p_2(x,\alpha)=\frac{\partial p_2(x,\alpha)}{\partial x}=0$  characterizes the
occurrence of a degenerate eigenvalue of ${\mathbbm J}({\boldsymbol\lambda})$. Inserting
the value $x=\hat{\jmath}= \frac{1}{2} (-\alpha -5)$  from (\ref{EC2a})
yields the critical value for $\alpha$
\begin{equation}\label{EC3}
 \alpha_c^{(1)}= -\frac{3}{5}
\end{equation}
that separates the collinear from the coplanar ground states.

\begin{figure}[t]
  \centering
    \includegraphics[width=1.0\linewidth]{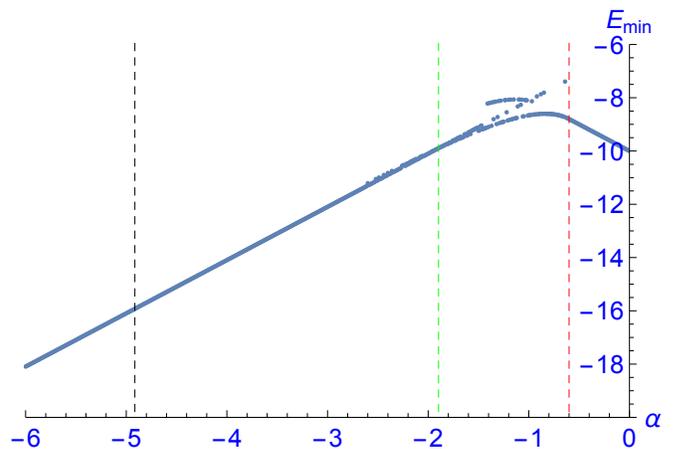}
  \caption
  {The numerically determined ground state energy $E_{min}$ of the almost regular cube as a function of the variable bond coupling parameter $\alpha$.
  Some spurious configurations yield energies above the minimal one.
  The critical values $\alpha_c^{(i)},\;i=1,2,3,$ are indicated by vertical dashed lines. At these values the nature of the ground state
  changes in a manner explained in the text.
  }
  \label{FIGGSC1}
\end{figure}

The closer investigation of coplanar ground states leads us to the limits of the analytical treatment of the problem,
even if we use computer algebraic methods. At any case numerical calculations have to serve as an auxiliary tool.
The main result is the occurrence of two further critical values
\begin{equation}\label{EC4}
  \alpha_c^{(2)}\approx -1.8963,\quad \mbox{and }  \alpha_c^{(3)}\approx -4.8208
\;,
\end{equation}
such that for $\alpha_c^{(2)}< \alpha <  \alpha_c^{(1)}$ and $\alpha <  \alpha_c^{(3)}$ we obtain
isolated ground states and for $\alpha_c^{(3)}< \alpha <  \alpha_c^{(2)}$ cooperative ones.

For a first orientation we numerically calculate the ground state energy $E_{min}$ for values $-6<\alpha<0$
of the variable bond strength, see Figure \ref{FIGGSC1}. We use a very simple algorithm that starts with a random spin configuration
and successively lowers its energy by correcting the single spins ${\mathbf s}_\mu$ according to (\ref{T3}).
The linear domain $-0.6<\alpha <0$ corresponds to the collinear ground state
${\mathbf s}_{coll}=(\uparrow\downarrow\downarrow\uparrow\uparrow\downarrow\uparrow\downarrow)$ and $E_{min}$ given by (\ref{EC2a}).
It may happen that the sketched numerical procedure does not yield an approximation of the true ground state,
but is trapped in the neighbourhood of a local energy minimum or pseudo-minimum.
This can be seen in Figure \ref{FIGGSC1} where a couple of connected points
lying nearly on a line is markedly above the energy minimum. We have not tried to get rid of these spurious ground states but
rather will use them as a hint to the position of certain cooperative states that will become ground states for
values of $\alpha_c^{(3)}<\alpha<\alpha_c^{(2)}$. The other spurious points forming a concave curve are neglected.

In order to analyze the isolated coplanar ground states we first note that for alternating
spin configurations ${\mathbf s}_\mu=-{\mathbf s}_{\mu+1},\;\mu=1,3,5,7$ the energy can be  written as
\begin{equation}\label{EC5}
E= \frac{1}{2}
\left(
\begin{array}{r}
 {\mathbf s}_1 \\
 -{\mathbf s}_1 \\
 {\mathbf s}_3 \\
 -{\mathbf s}_3 \\
{\mathbf s}_5 \\
 -{\mathbf s}_5 \\
 {\mathbf s}_7 \\
 -{\mathbf s}_7 \\
\end{array}
\right)^\top
\left(
\begin{array}{cccccccc}
 0 & 1 & \alpha  & 0 & 0 & 0 & 0 & 1 \\
 1 & 0 & 0 & \alpha  & 0 & 0 & 1 & 0 \\
 \alpha  & 0 & 0 & 0 & 1 & 0 & 1 & 0 \\
 0 & \alpha  & 0 & 0 & 0 & 1 & 0 & 1 \\
 0 & 0 & 1 & 0 & 0 & 1 & 0 & 1 \\
 0 & 0 & 0 & 1 & 1 & 0 & 1 & 0 \\
 0 & 1 & 1 & 0 & 0 & 1 & 0 & 0 \\
 1 & 0 & 0 & 1 & 1 & 0 & 0 & 0 \\
\end{array}
\right)
\left(
\begin{array}{r}
 {\mathbf s}_1 \\
 -{\mathbf s}_1 \\
 {\mathbf s}_3 \\
 -{\mathbf s}_3 \\
{\mathbf s}_5 \\
 -{\mathbf s}_5 \\
 {\mathbf s}_7 \\
 -{\mathbf s}_7 \\
\end{array}
\right)
\end{equation}
\begin{equation}\label{EC6}
 =\frac{1}{2}
\left(
\begin{array}{c}
  {\mathbf s}_1 \\
  {\mathbf s}_3 \\
  {\mathbf s}_5 \\
  {\mathbf s}_7 \\
\end{array}
\right)^\top
\left(
\begin{array}{rrrr}
 -2 & 2 \alpha  & 0 & -2 \\
 2 \alpha  & 0 & 2 & 2 \\
 0 & 2 & -2 & -2 \\
 -2 & 2 & -2 & 0 \\
\end{array}
\right)
\left(
\begin{array}{c}
  {\mathbf s}_1 \\
  {\mathbf s}_3 \\
  {\mathbf s}_5 \\
  {\mathbf s}_7 \\
\end{array}
\right)
\equiv\frac{1}{2} \tilde{\mathbf s}^\top \tilde{\mathbbm J} \tilde{\mathbf s}
\;.
\end{equation}
Hence we may use the $4\times 4$-matrix $\tilde{\mathbbm J}$ in (\ref{EC6}) as the undressed ${\mathbbm J}$-matrix of an auxiliary
$N=4$ spin system and apply the methods described in this paper. The unfamiliar diagonal terms should not bother us.

First we look for degenerate eigenvalues of $\tilde{\mathbbm J}({\boldsymbol\lambda})$ and hence for solutions of
\begin{equation}\label{EC7}
0=p(x,{\boldsymbol\lambda})=\frac{\partial p(x,{\boldsymbol\lambda})}{\partial x}=\frac{\partial p(x,{\boldsymbol\lambda})}{\partial \lambda_i}
,\quad i=1,2,3
\;,
\end{equation}
where $p(x,{\boldsymbol\lambda})$ is the characteristic polynomial
$p(x,{\boldsymbol\lambda})=\det\left( \tilde{\mathbbm J}({\boldsymbol\lambda})-x\,{\mathbbm 1}\right)$.
The solutions of (\ref{EC7}) depend on a free parameter and can be written in the form
$\lambda_i=L_i(x),\; i=1,2,3$. This is in accordance with the dimension $L-2=1$ of the variety
$j_{min}^{(2)}({\boldsymbol\lambda})$ due to the rules of avoided level crossing, see Section \ref{sec:T}.

\begin{figure}[t]
  \centering
    \includegraphics[width=1.0\linewidth]{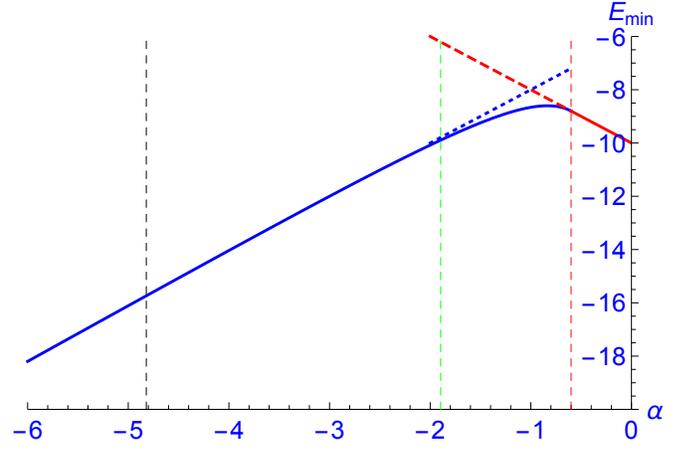}
  \caption
  { The analytically determined ground state energy $E_{iso}$ of the almost regular cube as a function of the variable bond coupling parameter $\alpha$
    for isolated coplanar ground states (blue curve).  For $\alpha\to -\infty$ the ground state energy is asymptotically given by
    $E_{asy}=2\alpha-6$ (dashed blue line).
    As in Figure \ref{FIGGSC1} the critical values $\alpha_c^{(i)},\;i=1,2,3,$ are indicated by
    vertical dashed lines. The energy of the collinear ground state according to (\ref{EC2a}) is represented by a red solid line in the domain
    $-\frac{3}{5} < \alpha <0$ and for smaller values of $\alpha$ by a dashed red line.
  }
  \label{FIGGSC2}
\end{figure}

Recall that $x$ represents the eigenvalue $j_{min}({\boldsymbol\lambda})$ of $\tilde{\mathbbm J}({\boldsymbol\lambda})$
and hence has to be chosen as the maximum of all values $x$ such that $L_i(x)$ is a smooth parameter representation.
It turns out that all $L_i(x)$ contain a square root of the form $\sqrt{f(x)}$ where $f(x)$ is the fourth order polynomial
\begin{eqnarray}
\nonumber
 f(x) &=& \left(16 \alpha ^2+8 \alpha +3\right) (\alpha -1)^4\\
\nonumber
&& -4 \left(8 \alpha ^2+1\right) (\alpha -1)^3 x \\
\nonumber
&& -4 (6 \alpha +1) (\alpha -1)^2    x^2\\
\label{EC8}
   && +16 \alpha  \left(2 \alpha ^2-\alpha +1\right) x^3-16 \alpha ^2 x^4
\;.
\end{eqnarray}
Hence the maximal value $\hat{\jmath}$ is given by a suitable zero of (\ref{EC8}) that
can be calculated analytically but is too complex to be given here. The corresponding
energy $E_{iso}(\alpha)= \frac{N}{2}\hat{\jmath} = 2 \hat{\jmath}$ is plotted in Figure \ref{FIGGSC2}.
This result is already very close to the numerical one, see Figure \ref{FIGGSC1}, but we will see that
for $\alpha_c^{(3)} < \alpha < \alpha_c^{(2)}$ the cooperative ground states will yield a smaller energy.

The collinear state ${\mathbf s}_{asy}=(\uparrow\downarrow\uparrow\downarrow\downarrow\uparrow\downarrow\uparrow)$
yields the energy $E_{asy}=2\alpha-6$ that is asymptotically assumed for $\alpha\to-\infty$, see Figure \ref{FIGGSC2}.

\begin{figure}[t]
  \centering
    \includegraphics[width=0.7\linewidth]{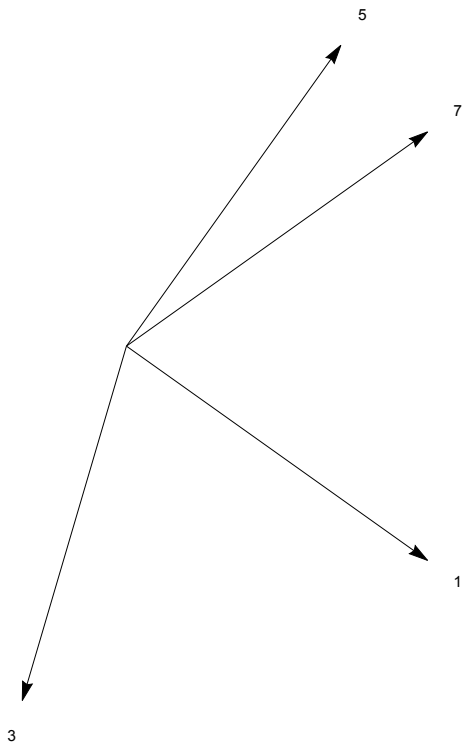}
  \caption
  {The analytically determined isolated ground state (\ref{EC9a}) - (\ref{EC9c})
  of the almost regular cube for the value of the variable bond coupling parameter $\alpha=-1$.
  Only spin vectors with odd numbers are shown, the even ones are obtained by ${\mathbf s}_{\mu+1}=-{\mathbf s}_{\mu},\;\mu=1,3,5,7$.
     }
  \label{FIGGSC3}
\end{figure}

The isolated ground state configurations corresponding to the minimal energy can also be analytically determined but only for
chosen values of $\alpha$. As an example we display the ground state configuration for $\alpha=-1$, see Figure \ref{FIGGSC3}.
This example is remarkable in so far as the ground state configuration can be characterized by a single angle:
\begin{eqnarray}\label{EC9a}
 && {\mathbf s}_1\cdot  {\mathbf s}_5=0
\quad \mbox{and}\\
\label{EC9b}
&&{\mathbf s}_1\cdot  {\mathbf s}_3=
{\mathbf s}_1\cdot  {\mathbf s}_7=\\
\label{EC9c}
&&
\frac{1}{4} \left(1-\sqrt{5}+\sqrt{2 \left(1+\sqrt{5}\right)}\right)
= 0.326993\ldots
\end{eqnarray}

Next we consider possible cooperative ground states.
There are no principal obstacles to obtaining closed solutions
depending on the parameter $\alpha$, but due to the practical
limitations of storage and computing times it is only
possible to  solve the problem for fixed values of $\alpha$.
Even with this restriction the intermediate  and final results
are too complex to be explicitly given.

\begin{figure}[t]
  \centering
    \includegraphics[width=1.0\linewidth]{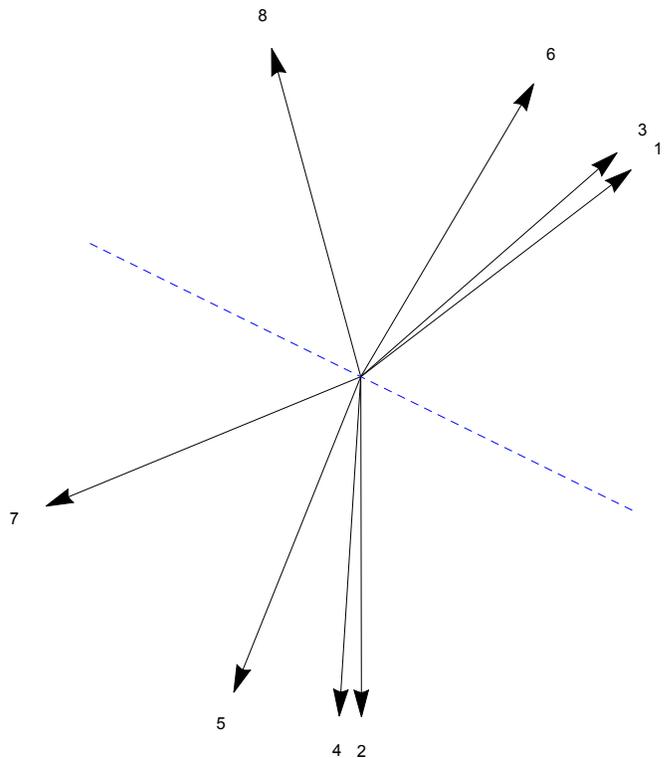}
  \caption
  {A numerically determined cooperative ground state of the almost regular cube for the value of the variable bond coupling parameter $\alpha=-2$.
   The symmetry axis is shown as a blue, dashed line.
  }
  \label{FIGCGS1}
\end{figure}

We return to the $N=8$ spin system and the dressed ${\mathbbm J}$-matrix (\ref{EC1}).
In this case the equations analogous to (\ref{EC7}) have solutions of the form $\lambda_i=K_i(x,\lambda_3),\;i=1,2,$ corresponding to a
two-dimensional intersection of eigenvalue varieties belonging to different eigenspaces of the symmetry $\Pi$.
This is in accordance with the dimension $L-1=2$ of the variety $j_{min}^{(2)}({\boldsymbol\lambda})$ due to
symmetry-allowed level crossing, see Section \ref{sec:T}.

In the following we consider the
example $\alpha=-2$ that is, however, typical for other integer or half integer values of $\alpha$. The function $\lambda_2=K_2(x,\lambda_3)$ has the form
of a certain double root of a polynomial $P_2(x,\lambda_2,\lambda_3)$. The corresponding equations
$0=P_2(x,\lambda_2,\lambda_3)=\frac{\partial P_2(x,\lambda_2,\lambda_3)}{\partial \lambda_2}$ can be used to eliminate $\lambda_2$
with the result $Q(x,\lambda_3)=0$, where $Q$ is the resultant of $P_2$ and its derivative w.~r.~t.~$\lambda_2$. Actually $Q$
is a polynomial in $x$ and $\lambda_3$ of degree $40$.
This step thus reduces the  intersection of the appropriate eigenvalue varieties of
${\mathbbm J}({\boldsymbol\lambda})$ to a curve of dimension one.
Since $x$ must be maximal within this curve we have the additional equation $\frac{\partial Q(x,\lambda_3)}{\partial \lambda_3}=0$.
Together with $Q(x,\lambda_3)=0$ we can numerically solve for $x$ and obtain $E_{min}= 4 x = -10.1009\ldots$.
Note that the essential use of numerics is confined to the last step of finding roots of polynomial equations.
A typical cooperative ground state for $\alpha=-2$ with the same symmetry as in Figure \ref{FIGGS1} is shown in Figure \ref{FIGCGS1}.

Analogously, we have obtained minimal energies $E_{coop}(\alpha)$ for the values $\alpha=-1,-1.5,\ldots,-6$, see Table \ref{tab1},
where also the values of $E_{iso}(\alpha)$ have been displayed.
The integer value $E_{iso}=-12$ for $\alpha=-3$ can be explained by the fact that here the isolated coplanar state degenerates
into the collinear state ${\mathbf s}_{asy}=(\uparrow\downarrow\uparrow\downarrow\downarrow\uparrow\downarrow\uparrow)$ with energy
$E(\alpha)=2\alpha-6$.

\begin{table}
\caption{{\label{tab1}}Table of energies $E_{coop}$ and $E_{iso}$ of possible cooperative/isolated ground states for different values of $\alpha$.
The ground state energies are shown in bold.}

\begin{center}\begin{tabular}{|c|c|c|}\hline
$\alpha$ &  $E_{coop}$ &  $E_{iso}$ \\ \hline\hline
$-1$&$-8.10796$& $\mathbf{-8.66038}$\\
\hline
$-1.5$&$-9.10359$ & $\mathbf{-9.24214}$\\
\hline
$-2$ & $\mathbf{-10.1009}$ & $-10.0787$\\
\hline
$-2.5$ & $\mathbf{-11.0992}$ & $-11.0154$\\
\hline
$-3$ & $\mathbf{-12.0979}$ & $-12.0000$\\
\hline
$-3.5$ & $\mathbf{-13.0969}$ & $-13.0105$\\
\hline
$-4$ & $\mathbf{-14.0962}$ & $-14.0359$\\
\hline
$-4.5$ & $\mathbf{-15.0956}$ & $-15.0703$\\
\hline
$-5$ & $-16.0951$ & $\mathbf{-16.1102}$\\
\hline
$-5.5$ & $-17.0947$ & $\mathbf{-17.1536}$\\
\hline
$-6$ & $-18.0943$ & $\mathbf{-18.1992}$\\
\hline
\end{tabular}
\end{center}
\end{table}

\begin{figure}[t]
  \centering
    \includegraphics[width=1.0\linewidth]{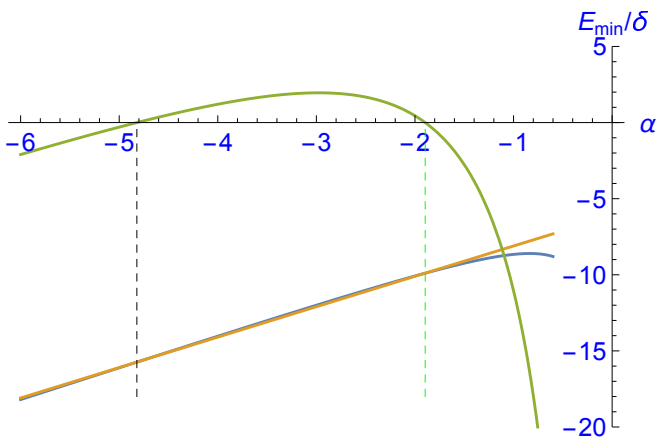}
  \caption
  {The two curves $E_{iso}(\alpha)$ (blue curve) and $E_{coop}(\alpha)$ (orange curve) and their intersections at
  $\alpha_c^{(2)}\approx -1.8963$ and $\alpha_c^{(3)}\approx -4.8208$ indicated by vertical dashed lines.
  Since the two curves can hardly be distinguished we have also shown
  the magnified difference $\delta \equiv 20(E_{iso}(\alpha)-E_{coop}(\alpha))$ (green curve).
   }
  \label{FIGCRIT}
\end{figure}

A cubic fit of the almost linear function $E_{coop}(\alpha)$ intersects the analytically determined function
$E_{iso}(\alpha)$ at the points $\alpha_c^{(2)}\approx -1.8963$ and $\alpha_c^{(3)}\approx -4.8208$, see Figure \ref{FIGCRIT},
such that for $\alpha_c^{(3)}< \alpha < \alpha_c^{(2)}$ the cooperative states represent the true ground states.
The phase transition at  $\alpha_c^{(3)}\approx -4.8208$ would be hardly detectable by purely numerical methods.

%%%%%%%%%%%%%%%%%%%%%%%%%%%%%%%%%%%%%%%%%%%%%%%%%%%%%%%%%%%%%%%%%%%%%%%%%%%%%%%%%%%%%%%%%%%%%%%%%%%%%%%%%%%%%%%%%%%%%%%%%%%%%%%%%%%%%%%%%%
\section{Summary and Outlook}\label{sec:SO}
%%%%%%%%%%%%%%%%%%%%%%%%%%%%%%%%%%%%%%%%%%%%%%%%%%%%%%%%%%%%%%%%%%%%%%%%%%%%%%%%%%%%%%%%%%%%%%%%%%%%%%%%%%%%%%%%%%%%%%%%%%%%%%%%%%%%%%%%%

We have revisited the theory of ground states published three years ago and added the novel aspect of the isolated/cooperative distinction
for ground states in the case of an involutary symmetry. This distinction has been illustrated by three examples, which also
have other interesting features: The isosceles triangle, the square with a diagonal bond and the almost regular cube.
All examples possess a variable bond parameter $\alpha$, and show a phase transition between
collinear and coplanar ground states at some critical values of $\alpha=\alpha_c$.
In all cases the critical values lie beyond the ``trivial domain" where the
collinear ground state minimizes the energy of every individual bond.
Thus we have a coexistence of competing interactions and collinear
ground states for
\begin{itemize}
  \item the triangle and $0< \alpha < 1/2$,
  \item the square and $0< \alpha < 3/2$, and
  \item the cube and $-3/5 < \alpha < 0$.
\end{itemize}
Moreover, the gauge parameters of the cube satisfy the remarkable identity
$\lambda_1=\lambda_2=-\lambda_3=-\lambda_4$ for the collinear ground state domain $-3/5=\alpha_c < \alpha$.
All examples possess cooperative coplanar ground states for suitable values of $\alpha$; the cube has additionally
isolated coplanar ground states and phase transitions between these two types at certain further critical values
of $\alpha$.

A natural generalization of these studies would consist in considering a larger group ${\sf G}$ of symmetries.
If ${\sf G}$ is generated by $d$ commuting involutary symmetries the generalization appears to be straight forward.
Instead of two orthogonal subspaces with different parity we would have $2^d$ ones and a corresponding variety of different
isolated and cooperative coplanar ground states.
A further generalization, however, would be faced with the following complication: Any non-involutary symmetry would possess complex eigenvalues
of the form $e^{{\sf i}\,\alpha}$ with $\alpha\neq 0, \pi$ and the corresponding complex eigenspaces.
Since the eigenspaces ${\mathcal W}_{min}(\hat{\boldsymbol\lambda})$ considered for the ground state problem have to be real
we cannot simply transfer the results of the present paper to the general case of a symmetry group.

Another possible generalization would consist of extending the
isolated/cooperative distinction and the
application of the avoided level crossing rules to three-dimensional
ground states.

%%%%%%%%%%%%%%%%%%%%%%%%%%%%%%%%%%%%%%%%%%%%%%%%%%%%%%%%%%%%%%%%%%%%%%%%%%%%%%%%%%%%%%%%%%%%%%%%%%%%%%%%%%%%%%%%%%%%%%%%%%%%%%%%%%%%%%%%%%
%\section*{Acknowledgment}
%%%%%%%%%%%%%%%%%%%%%%%%%%%%%%%%%%%%%%%%%%%%%%%%%%%%%%%%%%%%%%%%%%%%%%%%%%%%%%%%%%%%%%%%%%%%%%%%%%%%%%%%%%%%%%%%%%%%%%%%%%%%%%%%%%%%%%%%%

%\section*{References}

\end{document}